\documentclass{eas}
\usepackage{graphicx}
\begin{document}

\title{Past singularities in phantom theories} 
\author{Leonardo Fern\'andez-Jambrina}\address{E.T.S.I. 
Navales\\Universidad Polit\'ecnica de Madrid \\Arco de la Victoria 
s/n\\ 28040-Madrid, Spain}
\begin{abstract}
FLRW models filled with just dark energy are shown to have a finite
past, since causal geodesics cannot be extended beyond a certain
proper time. It is shown that curvature measured along causal
geodesics becomes infinity on travelling to the past, though 
curvature scalars tend to zero. Furthermore the time measured by 
free-falling observers from coincidence time to Big Rip is shown to 
be as short as wished by increasing their linear momentum.
\end{abstract}
\maketitle
\section{Introduction}

The analysis of several sources of observational data, supernovae type
Ia, redshift of distant objects, temperature fluctuations of
background radiation (cfr.  Padmanabhan \cite{de} for a nice review)
has lead to accept that our universe is acceleratedly expanding
nowadays.

This unexpected feature, which cannot be explained by means of an
standard matter content for the universe, has motivated lots of
theoretical work, such as a dark energy component in the
energy-momentum tensor or modifications of the theory of 
gravitation. Violation of energy conditions due to these new 
components have given rise to consider new types of singularities, 
different from the well-known Big Bang and Big Crunch. Beyond the 
so-called phantom divide, $w<-1$, accelerated expansion ends up at a 
Big Rip (Caldwell \cite{caldwell}). Below  the phantom divide, other
atypical events may come 
up, such as sudden singularities (Barrow \cite{sudden}).

Usually the analysis of these singularities is made by just tracking
where the curvature scalars blow up.  For instance, a classification
of singularities is provided (Nojiri and Odintsov \cite{classodi})
depending on which scalars are singular or not.  However the use of
more poweful tools, such as geodesic analysis, has proven quite
successful to discover new features of the singularities.
Furthermore, this is the natural framework to talk about singularities
(Hawking and Ellis \cite{HE}).

\section{FLRW models}

We consider homogeneous and isotropic universes endowed with a
Friedmann-Lema\^{i}tre-Robertson-Walker metric 
\begin{eqnarray*}
   ds^2=-dt^2+a^2(t)\left\{f^2(r)dr^2+ r^2d\Omega^2\right\},
    \quad f^2(r)=\frac{1}{1-kr^2},\quad 
    k=0,\pm1,\end{eqnarray*}
where the coordinates have the usual ranges. We are mostly interested 
in spatially-flat models, $k=0$, since observations point out this 
kind of universe.

Geodesics in these models are described by simple ordinary equations 
due to their high symmetry,
\begin{eqnarray}\label{geods}
    \dot t&=&\sqrt{\delta +\frac{P^2}{a^2(t)}},\label{geods1}\\\dot
    r&=&\pm\frac 
    {P}{a^2(t)f(r)}.\label{geods2}\end{eqnarray}
   
The dot stands for derivation with respect to proper time $\tau$, 
which allows parametrization of geodesics satisfying the unitary 
condition of the velocity,
\[\delta =\dot t^2-a^2(t)f^2(r) \dot r^2,\]where $\delta$ is one for 
timelike, zero for lightlike or minus one for spacelike geodesics. 
The constant $P$ is the specific linear momentum of the geodesic.

There are then three families of causal geodesics: lightlike 
geodesics ($\delta=0$), fluid worldlines ($\delta=1$, $P=0$) and 
timelike radial geodesics ($\delta=1$, $P\neq0$).

In order to perform calculations with a general scale factor, we only
need that $a(t)$ has a power expansion (Catto\"en and Visser
\cite{visser}) around the event at $t_{0}$, where a singularity is
deemed to appear,
\[a(t)=c_{0}|t-t_{0}|^{\eta_{0}}+c_{1}|t-t_{1}|^{\eta_{1}}+\cdots, \]
with real and ordered exponents, $\eta_{0}<\eta_{1}\cdots$, and
positive $c_{0}$.  Hence, the model behaves close to $t_{0}$ as a
power-law model of exponent $\eta_{0}$.  Accordingly, several types of
singular events may arise:

\begin{itemize}
\item  $\eta_{0}>0$: the scale factor vanishes at $t_{0}$ and
we have either a Big Bang or Big Crunch.

\item $\eta_{0}=0$: the scale factor is finite at $t_{0}$.  If
$a(t)$ is analytical, the event at $t_{0}$ is regular.
Otherwise a weak or sudden singularity appears.

\item $\eta_{0}<0$: the scale factor diverges at $t_{0}$ and
a Big Rip appears. 
\end{itemize}

The analysis of causal geodesics is easy to perform in this case 
and is shown in Table 1 (Fern\'andez-Jambrina and Lazkoz \cite{puiseux}).

\begin{table}\begin{center}\begin{tabular}{|c|c|c|c|c|c|}
	\hline
	${\eta_{0}}$ & ${\eta_{1}}$ & ${k}$ & $c_{0}$ &\textbf{Tipler} &
	\textbf{Kr\'olak}  \\
	\hline\hline
	$(-\infty,0)$ & $(\eta_{0},\infty)$ &   && 
	Strong & Strong  \\
	\cline{1-2} \cline{5-6}
	$0$ & $(0,1)$ &  && Weak & Strong  \\
	    \cline{2-2} \cline{5-6}
	     & $[1,\infty)$ & $0,\pm 1$  &$(0,\infty)$& Weak & Weak
\\
	     \cline{1-2} \cline{5-6}
	$(0,1)$ & $(\eta_{0},\infty)$ &   && Strong & Strong  \\
	\cline{1-3} \cline{5-6}
	$1$ & $(1,\infty)$ & $0,1$  && Strong & Strong  \\
	\cline{2-6}
	 & $(1,\infty)$ &  &$(0,1)\cup(1,\infty)$ & Strong & Strong  \\
	 \cline{2-2} \cline{4-6}
	 & $(1,3)$ & $-1$ & 1 & Weak & Strong  \\
	 \cline{2-2} \cline{5-6}
	 & $[3,\infty)$ &  &  &Weak & Weak  \\
    \hline
	$(1,\infty)$ & $(\eta_{0},\infty)$ & $0,\pm 1$  &$(0,\infty)$&
Strong & Strong \\
	\hline\end{tabular}\caption{Singularities in FLRW 
	models}\end{center}\end{table}

This table shows the first of the unusual features of sudden
singularities (Fer\-n\'an\-dez-Jambrina and Lazkoz \cite{flrw}): they are
weak according to Tipler (Tipler \cite{tipler}) and Kr\'olak's definitions
(Krolak \cite{krolak}) and therefore they do not exert infinite tidal forces
that might destroy extended bodies going through them.  In this sense,
they cannot be considered the final fate of a universe, since they can
be avoided by some observers.

Concerning the Big Rip, another issue unveiled by the survey of 
causal geodesics in FLRW models with an effective equation of state 
$p=w\rho$, $w\in(-5/3,-1)$ for late times is that lightlike geodesics 
need infinite proper time before meeting the Big Rip. That is, only 
massive objects see the Big Rip. In this sense, since photons do not 
reach this event, we could say that Big Rip may be considered quite 
an intimate experience\ldots{} with lights turned out! Since dark 
energy models with $w$ just below minus one have been shown to be 
compatible with observations, this features cannot be neglected as 
mathematical curiosities.

\section{What about an infinite $t_{0}$?}

The issue of singularities located at infinite time coordinate is not 
as pointless as it may seem at first glance. The example of photons 
close to a Big Rip shows us that a finite coordinate time lapse can 
be experienced as an infinite proper time lapse, in this case by 
massless particles. The relation between proper time and coordinate 
time is far from trivial.

First, we have to determine when causal geodesics reach infinite 
coordinate time in finite proper time. 

For lightlike geodesics we may integrate the geodesic equation 
(\ref{geods1}),
\[ \dot t=\frac{P}{a(t)}\Rightarrow 
\int_{t_{0}}^t a(t')\,dt'=P(\tau-\tau_{0}),\]
in order to attain a necessary and sufficient condition for reaching 
infinite coordinate time in finite proper time: integrability of the 
scale factor $a(t)$,
\begin{equation}\label{conda}\int_{t}^\infty a(t')\,dt'<\infty,\end{equation}
for sufficiently large $t$. 

The same condition is valid for radial timelike geodesics. On the 
contrary, fluid worldlines with $P=0$ are not to be considered since 
for them proper time is essentially coordinate time.

For a start, we may consider models which behave for large $t$ as a 
power-law model $a(t)\simeq ct^{\eta}$. For them the geodesic equation 
(\ref{geods1}) yields
\begin{eqnarray*}t\simeq\left\{\begin{array}{ll}\displaystyle\left\{\frac{(1+\eta) P}{c}\right\}^{1/(1+\eta)}
(\tau-\tau_{0})^{1/(1+\eta)} & \eta\neq-1,\\\\ \displaystyle
e^{P(\tau-\tau_{0})/c} & \eta=-1,\end{array}\right.\end{eqnarray*}
and obviously infinite $t$ is reached for finite $\tau_{0}$ if and only 
if $\eta<-1$.

This is not however the interesting case, since it involves models 
starting with a Big Rip at $t=0$. But if we consider models starting 
at $t=-\infty$ and ending up at a Big Rip at $t=0$ the same reasoning 
is valid. The origin of radial causal geodesics would be at a finite 
proper time.

At this point, two possibilities arise in order to explain the finite
origin of causal geodesics in phantom models: there is either a
singularity at $t=-\infty$ which explains the abrupt startpoint of
these geodesics instead of ranging from and infinite proper time or
the FLRW coordinate patch does not cover the whole universe and we
need to enlarge it.

This is done, for instance, with de Sitter universes in the
parametrization which is used for inflation, $k=0$,
$a(t)=e^{\sqrt{\Lambda/3}\,t}$, which fulfills condition (\ref{conda})
and therefore its radial geodesics reach $t=-\infty$ in finite proper
time.  However, in this case it is possible to extend the spacetime to
a larger one, $k=1$,
$a(T)=\sqrt{3/\Lambda}\cosh\left(\sqrt{\Lambda/3}\,T\right)$ with
another change of coordinates (Hawking and Ellis \cite{HE}) and hence
the singularity at $t=-\infty$ is only apparent and due to a bad 
choice of coordinates.

This option appears to be the correct for other universes since at
$t=-\infty$ all curvature scalars vanish as $1/t^2$ for power-law
models.

However if we reckon Ricci tensor components along radial 
lightlike geodesics, unexpected features arise,
\[u^t=\frac{P}{a},\quad u^r=\pm\frac{P}{fa^2},\quad
R_{ij}u^iu^j=2P^2\left(\frac{a'^2+k}{a^4}-\frac{a''}{a^3}\right)\simeq
\frac{2P^2\eta}{c^2t^{2(\eta+1)}}
+\frac{2kP^2}{c^4t^{4\eta}},\]
since the first term diverges at $t=-\infty$ for $\eta<-1$ regardless 
of the value of $k$. The same happens for radial timelike geodesics.

This unexpected result could only be avoided in models for which the 
curvature term cancels the first term, $a'^2+k=aa''$. But curiously 
the solutions of this equation are well known universes.

For $k=0$ we recover the aforementioned parametrization of de Sitter
universe, $a(t)=e^{\sqrt{\Lambda/3}\,t}$.  For $k=1$ we find another
parametrization of de Sitter universe,
$a(t)=\sqrt{3/\Lambda}\cosh\left(\sqrt{\Lambda/3}\,t\right)$.  And for
$k=-1$, also another parametrization of de Sitter spacetime arises,
$a(t)=\sqrt{3/\Lambda}\sinh\left(\sqrt{\Lambda/3}\,t\right)$ but also
anti-de Sitter spacetime,
$a(t)=\sqrt{3/\Lambda}\cos\left(\sqrt{\Lambda/3}\,t\right)$, but for a
choice of time coordinate origin.  Since only the one with $k=0$
decreases at $t=\pm\infty$, they do not affect our results.

Therefore we are led to conclude that universes which asymptotically
behave as power-law models with $\eta<-1$, $w\in(-5/3,-1)$, in spite
of having vanishing curvature scalars at $t=-\infty$, they possess
there a sort of directional singularity that affects only radial
geodesics, a p.p. curvature singularity (curvature singularity along a
parallelly transported basis) (Hawking and Ellis \cite{HE}). 
Furthermore, this singularity is strong (Tipler \cite{tipler}, 
Kr\'olak \cite{krolak}) and therefore cannot be overlooked.

Radial observers would start their trajectories at a singularity at
$t=-\infty$ and would end up at a Big Rip at $t=0$, though their life
span would be finite, as we have shown.  Photons would start their
paths at the singularity at $t=-\infty$ but they would never reach the
Big Rip in these models.

Summarizing these results and extending the conclusions to 
non-asymptotically power-law models, we have:

\begin{itemize}
    \item For scale factors growing or decreasing as $1/|t|$ or
    slower, radial causal geodesics reach $t=-\infty$ in infinite
    proper time.

    \item For scale factors decreasing faster than $1/|t|$, radial
    causal geodesics reach $t=-\infty$ in finite proper time, meeting
    a strong curvature singularity, except for de Sitter spacetime.
\end{itemize}

Cases with an ill-defined limit at $t=-\infty$ may be handled directly
with condition (\ref{conda}).

\section{Duration of the universe}

The features which have been shown about the past of phantom models
might seem useless since dark energy is a relevant component only for
the future of the universe from coincidence time on in order to
explain accelerated expansion, not for the past.  Nonetheless there
are implications for the models.

For instance, we may use our results to reckon the duration of a 
universe filled with a power-law phantom field as measured by an 
observer following a radial geodesic (\ref{geods1}),
\begin{eqnarray}T&=&\int_{-\infty}^0\frac{dt}{\sqrt{1+P^2/a^2(t)}}=
\int_{-\infty}^0\frac{dt}{\sqrt{1+P^2/c^2t^{2\eta}}}\nonumber\\&=&
\left(\frac{P}{c}\right)^{1/\eta}\int^{\infty}_0\frac{x^\eta dx}{\sqrt{1+x^{2\eta}}}
=\left(\frac{P}{c}\right)^{1/\eta} I,
\label{P}\end{eqnarray}
which is finite again for $\eta<-1$.

But most surprising is that this quantity can be made as small as 
wanted by increasing the total linear momentum $P$ of the traveller. 
It may be argued that this calculation is useless since we cannot 
extend the validity of the phantom model towards the past, but this 
does not invalidates the result. We may calculate the time span 
measured by the observer from coincidence time to the Big Rip and 
reach the same conclusion: observers may reduce the life of the 
universe by increasing their linear momentum. We have always known 
that driving fast could be dangerous, but not to this extent! 
Since a negative exponent $\eta$ is necessary in (\ref{P}) for this 
result, it is clear that this behaviour is exclusive of phantom 
models with $w<-1$ for large times, that is, below the phantom divide.

So far we have shown that phantom fields produce not only a Big Rip 
but a directional singularity in the past, that massless particles do 
not experience a Big Rip and that the duration of the universe may be 
shortened \emph{ad libitum} by increasing the linear momentum of the 
observer. There are other exotic features inherent to phantom models 
(Fern\'andez-Jam\-bri\-na \cite{plb}), especially of those which are 
closer to the phantom divide, but these should be enough to show the 
unexpected behaviour that one encounters on trying to model dark 
energy with phantom fields.

\section*{Acknowledgments}This work is supported by the Spanish
Ministry of Education and Science research grant FIS-2005-05198.  The
author wishes to thank R. Lazkoz, J.M.M. Senovilla and R. Vera for
valuable discussions and the University of the Basque Country for
their hospitality.  The author also wants to thank the organizers of
the 2007 Spanish Relativity Meeting for a superb conference.


\end{document}